\documentclass{eas}
\usepackage{graphicx}
\usepackage{natbib}
\begin{document}

%%-----------------------------
%%      the top matter
%%-----------------------------
\title{Star Formation Patterns and Hierarchies}
\author{Bruce G. Elmegreen}\address{IBM T. J. Watson Research Center, 1101 Kitchawan Road, Yorktown
Heights, New York 10598 USA, bge@us.ibm.com}
\begin{abstract}
Star formation occurs in hierarchical patterns in both space and time.
Galaxies form large regions on the scale of the interstellar Jeans
length and these large regions apparently fragment into giant molecular
clouds and cloud cores in a sequence of decreasing size and increasing
density. Young stars follow this pattern, producing star complexes on
the largest scales, OB associations on smaller scales, and so on down
to star clusters and individual stars. Inside each scale and during the
lifetime of the cloud on that scale, smaller regions come and go in a
hierarchy of time. As a result, cluster positions are correlated with
power law functions, and so are their ages. At the lowest level in the
hierarchy, clusters are observed to form in pairs. For any hierarchy
like this, the efficiency is automatically highest in the densest
regions. This high efficiency promotes bound cluster formation. Also
for any hierarchy, the mass function of the components is a power law
with a slope of around $-2$, as observed for clusters.
\end{abstract}
\maketitle

\section{Introduction}

Local Open Clusters in the compilation by \cite{piskunov06} seem at
first to have a random distribution in the galactic plane, with the
local spiral arms barely visible and no obvious age gradients or
patterns. They have been known to be grouped into star complexes
\citep{ef95} and moving groups \citep{eggen89} for a long time, but
there has been little other patterning recognized. Now this is
beginning to change. The groupings and complexes are better mapped
using new velocity and distance information. \cite{piskunov06} and
\cite{khar05} catalogued ``Open Cluster Complexes,'' in which many
clusters have similar positions, velocities and ages inside each
complex. For example, one is in the Hyades region and another is in
Perseus-Auriga. Perseus-Auriga surrounds the Sun and lies in the
galactic plane over a region 1 kpc in size with a log(age) between 8.3
and 8.6, in years. Gould's Belt is another Open Cluster Complex. It has
a log age less than 7.9 and lies in a thin plane tilted to the main
galactic disk by an angle of 20$^\circ$ surrounding the Sun.

\cite{ff08} identified five Open Cluster Complexes from the positions
and velocities of clusters within 2.5 kpc of Sun. These are:
Scutum-Sagittarius at a galactic longitude of $l=12^\circ$ and a
distance of 1300 pc, Cygnus at $l=75^\circ$ and 1400 pc,
Cassiopeia-Perseus at $l=132^\circ$ and 2000 pc, Orion at $l=200^\circ$
and 500 pc, and Centaurus-Carina at $l=295^\circ$ and 2000 pc. These
authors suggest that Open Cluster Complexes are fragments from common
gas clouds. Within their limiting distance of 2.5 kpc, the total gas
mass in the Milky Way is $\sim5\times10^7\;M_\odot$, considering a disk
thickness of 300 pc and an average density of 1 cm$^{-3}$. This means
that each of the 5 giant gas clouds that made these Open Cluster
Complexes had a mass of $\sim10^7\;M_\odot$. This is the Jeans mass in
the galactic disk ($\sim\sigma^4/[G^2\Sigma_{\rm gas}]$ for dispersion
$\sigma$ and mass column density $\Sigma_{\rm gas}$), as discussed in
Lecture 1.  Open Cluster Complexes could be the remnants of star
formation in giant clouds formed by gravitational instabilities in the
Milky Way gas layer.

\cite{elias09} studied Gould's Belt using the Catalogue of Open Cluster
Data \citep{khar05}.  They found an interesting correlation that the
cluster fraction is large for the Orion OB association region and small
for the Sco-Cen association. The cluster fraction is the ratio of the
stellar mass that forms in bound clusters to the total stellar mass
that forms at the same time. The rest of the stars form in unbound
groups and associations. There is a gradient in the young cluster (age
$<10$ Myr) fraction of star formation and in the cluster density over
the 700 pc distance separating these two associations. This suggests
that star formation prefers clusters when the pressure is high, as in
Orion, which is a more active region than Sco-Cen. High pressure could
be a factor in bound cluster formation if high-pressure cores are more
difficult to disrupt and their star formation efficiencies end up
higher when star formation stops. High pressure also corresponds to a
broader density probability distribution function, and so a higher mass
fraction of gas exceeding the critical efficiency for bound cluster
formation (Sect. \ref{sect:uc}).

This lecture reviews interstellar and stellar hierarchical structure,
which gives patterns in the positions and ages of young stars and
clusters. Related to this is the formation of the bound clusters
themselves, and the cluster mass function. A more complete review of
this topic is in \cite{e10}.

\section{Galactic Scale}

The hierarchy of star formation begins on the scale of Jeans-mass cloud
complexes. Giant molecular clouds (GMCs) form by molecular line
shielding and condensation inside these giant clouds, and star
complexes build up from the combined star formation \citep{ee83}. Each
GMC makes a single OB association at any one time. Hierarchical
structure has been known to be important in star-forming regions for a
long time \citep[e.g.,][]{larson81, fg87}. Early reviews of large-scale
hierarchical structure are in \cite{scalo85,scalo90}. Interstellar
hierarchies have also been thought to have a possible role in the
stellar initial mass function
\citep[e.g.,][]{larson73,larson82,larson91}.

\begin{figure}[b]
% \vspace*{-2.0 cm}
\begin{center}
 \includegraphics[width=5.in]{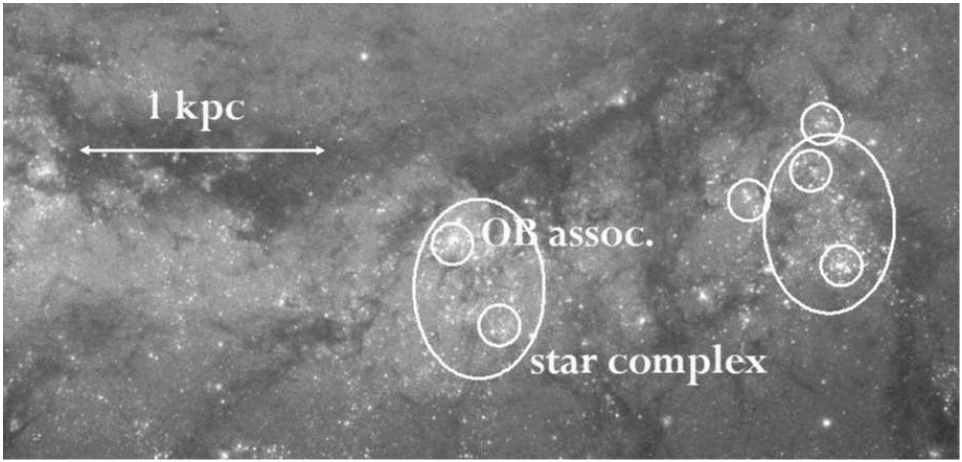}
% \vspace*{-1.0 cm}
\caption{The Southern part of the inner spiral arm of M51, showing star
formation on a variety of scales, with OB associations inside star
complexes and gas structures all around. The dust lane is broken up
into giant cloud complexes that contain $10^7\;M_\odot$.}\label{f1}
\end{center}
\end{figure}

The nearby galaxy M33 has a clear pattern of giant HI clouds, with
masses of $10^6-10^7\;M_\odot$, containing most of the GMCs and CO
emission \citep{engargiola03}. Giant star complexes occur in these
regions \citep{ivanov05}, often extending beyond the HI clouds because
of stellar drift. The high-definition image of M51 made by the ACS
camera on the Hubble Space Telescope shows exquisite examples of
stellar clustering on a wide variety of scales, with similar patterns
of clustering for dust clouds, which are the GMCs (Fig. \ref{f1}).
Clearly present are giant clouds (1 kpc large with
$M\sim10^7\;M_\odot$) that are condensations in spiral arm dust lanes,
star formation inside these clouds with no noticeable time delay after
the spiral shock, and scattered star formation downstream. The
downstream activity has the form of lingering star formation in cloud
pieces that come from the disassembly of spiral arm clouds, in addition
to triggered star formation in shells and comet-shaped clouds that are
also made from the debris of spiral arm clouds (see Lectures 2 and 4).

Star clusters in M51 observed with HST have been studied by
\cite{scheep09}. The overall distribution of clusters in the M51 disk
shows no obvious correlations or structures, aside from spiral arms.
But autocorrelation functions for three separate age bins show that the
youngest sample is well correlated: it is hierarchical with a fractal
dimension of $\sim1.6$. This means that there are clusters inside
cluster pairs and triplets, that are inside clusters complexes and so
on, up to $\sim1$ kpc. Clusters in the Antennae galaxy are also
auto-correlated out to $\sim1$ kpc scales \citep{zfw01}.

\cite{sanchez08} surveyed the positions of HII regions in several
galaxies. They found that the fractal dimension, $D_{\rm c}$, of the
distribution of HII regions decreases with increasing HII region
brightness.  For NGC 6946, $D_{\rm c} = 1.64$ for high-brightness HII
regions, $D_{\rm c} = 1.82$ for medium-brightness, and $D_{\rm c} =
1.79$ for low-brightness.  They also found that among galaxies with
more than 200 HII regions, the fractal dimension decreases slightly
with decreasing galaxy brightness.

The fractal dimension is the ratio of the log of the number $N$ of
substructures in a region to the log of the relative size $S$ of these
substructures. If we imagine a square divided into $3\times3$
subsquares, which are each divided into $3\times3$ more subsquares, and
so on, then the size ratio is $S=3$ for each level. If 6 of these
subsquares actually contain an object like an HII region (so the
angular filling factor is 6/9), then $N=6$ and the fractal dimension is
$\log 6 / \log 3= 1.63$. If all 9 regions contain substructure, then
the fractal dimension would be $\log 9 / \log 3=2$, which is the
physical dimension of the region, viewed in a 2-dimensional projection
on the sky.  Thus a low fractal dimension means a small filling factor
for each substructure in a hierarchy of substructures. If the brightest
HII regions have the smallest fractal dimension in a galaxy, then this
means that the brightest HII regions are more clustered together into a
smaller fraction of the projected area. This greater clustering is also
evident from maps of the HII region positions as a function of
brightness \citep{sanchez08}. The brightest HII regions tend to be
clustered tightly around the spiral arms. Similarly, fainter galaxies
have more tightly clustered HII regions than brighter galaxies.

The size distributions of star-forming regions can also be found by
box-counting. \cite{eec06} blurred an HST/ACS image of the galaxy NGC
628 in successive stages and counted all of the optical sources at each
stage with the software package {\it SExtractor}. The cumulative size
distribution of structures, which are mostly star-forming regions, was
a power law with power 2.5 for all available passbands, B, V, and I,
i.e., $n(>R)dR\propto R^{-2.5} dR$ for size $R$. The H$\alpha$ band had
a slightly shallower power.  They compared this distribution with the
distribution of structures in a projected 3D model galaxy made as a
fractal Brownian motion density field. They got good agreement when the
power spectrum for the model equalled the 3D power spectrum of
Kolmogorov turbulence, which has a slope of 3.66. Other power spectra
gave either too little clumpiness of the structures (lower $n[>R]$
slope) or too much clumpiness (higher slope).

Azimuthal intensity profiles of optical light from galaxies have
power-law power spectra like turbulence too. \cite{eel03} showed that
young stars and dust clouds in NGC 5055 and M81 have the same
scale-free distribution as HI gas in the LMC \citep{eks01}, both of
which have a Kolmogorov power spectrum of structure. For azimuthal
scans, the power spectrum slope was $\sim-5/3$ in all of these cases.
\cite{block09} made power spectra of Spitzer images of galaxies. They
included M33, which is patchy at 8$\mu$m where PAH emission dominates,
and relatively smooth at $3.6\mu$m and $4.5\mu$m where the old stellar
structure dominates. The power spectra showed this difference too: the
slopes were the same as those of pure noise power spectra for the
stellar images, and about the same as Kolmogorov turbulence for the PAH
images. The grand-design galaxy M81 had the same pattern. Block et al.
also reconstructed images of the galaxies over the range of Fourier
components that gave the power-law power spectrum. These images
highlight the resolved hierarchical parts of the galaxies. These parts
are primarily star complexes and young stellar streams.

The central regions of some galaxies show highly structured dust clouds
in HST images. The central disk in the ACS image of M51 shows this, for
example. In the central regions of these galaxies, there are large
shear rates, strong tidal-forces, sub-threshold column densities,
strong radiation fields, and lots of holes and filaments in the dust.
The origin of the holes is not known, although it is probably a
combination of radiation pressure, stellar winds, and turbulence.
Irregular dust in the center of NGC 4736 was studied by \cite{eee02}
using two techniques. One used unsharp mask images, which are
differences between two smoothed images made with different Gaussian
smoothing functions. Unsharp masked images show structure within the
range of scales given by the smoothing functions. They also made power
spectra of azimuthal scans. The power spectra were found to be power
laws with a slope of around $-5/3$, the same as the slope for HI in the
LMC and optical emission in NGC 5055 and M81. A possible explanation
for the power-law dust structure in galactic nuclei is that this is a
network of turbulent acoustic waves that have steepened into shocks as
they move toward the center \citep{montenegro99}.

\begin{figure}[b]
% \vspace*{-2.0 cm}
\begin{center}
 \includegraphics[width=5.in]{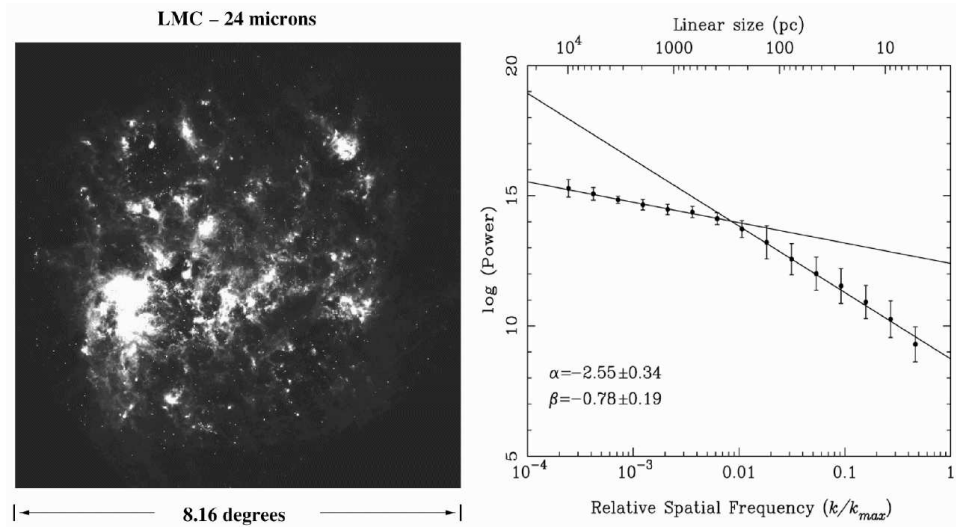}
% \vspace*{-1.0 cm}
\caption{(Left) An image of the LMC at 24$\mu$m from the Spitzer Space
Telescope. (Right) The 2D power spectrum of this image, showing two
power-law regions. The region with high slope at large spatial
frequency $k$ is presumably 3D turbulence inside the thickness of the
disk, and the region with low slope at small $k$ is presumably from 2D
turbulence and other motions on larger scales.  The break in the slope
defines the scale of the disk thickness
\citep[from][]{block10}.}\label{f2}
\end{center}
\end{figure}

\cite{block10} made power spectra of the Spitzer images of the Large
Magellanic Cloud at $160\mu$, $70\mu$, and $24\mu$ (see Figure
\ref{f2}). Again the power spectra are power laws, but now the power
laws have breaks in the middle, as found previously in HI images of the
LMC \citep{eks01}.  These breaks appear to occur at a wavenumber that
is comparable to the inverse of the disk line-of-sight thickness. On
scales smaller than the break, the turbulence is 3D and has the steep
power spectrum expected for 3D, and on larger scales the turbulence is
2D and has the expected shallower spectrum. The LMC is close enough
that each part of the power spectrum spans nearly two orders of
magnitude in scale. The slopes get shallower as the wavelength of the
observation decreases, so there is more small-scale structure in the
hotter dust emission.

Power-law power spectra in HI emission from several other galaxies were
studied by Dutta and collaborators. \cite{dut08} obtained a power
spectrum slope of $-1.7$ covering a factor of 10 in scale for NGC 628.
\cite{dut09a} found two slopes in NGC 1058 with a steepening from $-1$
to $-2.5$ at an extrapolated disk thickness of 490 pc (although their
spatial resolution did not resolve this length). H$\alpha$ and HI power
spectra of dwarf galaxies showed single power laws
\citep{willett05,begum06,dut09b}, as did the HI emission
\citep{stanimirovic99} and dust emission \citep{stanimirovic00} of the
Small Magellanic Cloud. The SMC has an interesting contrast to the LMC,
both of which are close enough to make a power spectrum over a wide
range of spatial scales. The LMC has a two component power spectrum,
but the SMC has only a single power-law slope. The difference could be
because the line-of-sight depth is about as long as the transverse size
for the SMC, while the line-of-sight depth is much smaller than the
transverse size for the LMC.

\section{Time-Space Correlations}

In addition to being correlated in space, clusters are also correlated
in time. \cite{efe98} found that the age difference between clusters in
the LMC increases with the spatial separation as a power-law, age
$\propto$ separation$^{1/2}$. \cite{eef96} found the same
age-separation correlation for Cepheid variables in the LMC.
\cite{ff09a} showed that this correlation also applies to clusters in
the solar neighborhood. In both cases, the correlation is strongest for
young clusters with a separation less than $\sim1$ kpc, and it goes
away for older clusters. Presumably the young clusters follow the
correlated structure that the gas had when the clusters formed.
Clusters form faster in regions with higher densities out to a kpc or
so, which is probably the ISM Jeans length. This means that small
star-forming regions (e.g., cluster cores) come and go during the life
of a larger star-forming region (an OB association), and then the
larger regions come and go during the life of an even larger region (a
star complex). Eventually, the clusters, associations and complexes
disperse when they age, taking more random positions after $\sim100$
Myr.  The correlation is about the same as the size-linewidth relation
for molecular clouds \citep{larson81}, considering that the ratio of
the size to the linewidth is a timescale.

Correlated star formation implies that some clusters should form in
pairs. Cluster pairs were discovered in the LMC by \cite{bhatia88} and
in the SMC by \cite{hb90}. An example is the pair NGC 3293 and NGC 3324
near eta Carinae. \cite{ff09b} studied these and other pairs. For NGC
3292/3324, the clusters are apparently weakly interacting and the age
difference is 4.7 My. NGC 659 and NGC 663 are also weakly interacting
and the age difference is 19.1 My. \cite{dieball02} determined the
distribution function of the number of cluster members per cluster
group in the LMC. They found a statistical excess of clusters in pairs
compared to the expectation from random groupings.

\section{Correlated Star Positions}

Individual young stars are correlated in position too. \cite{gomez93}
studied the 2-point correlation function for stars in the local
star-forming region Taurus. For 121 young stars, there was a power law
distribution of the number of stars as a function of their separation,
which means that the stars are hierarchically correlated. (A similar
correlation might arise from an isothermal distribution of stars
without any hierarchical structure, as in a relaxed star cluster, but
the Taurus region is not like this.) The correlation in Taurus extended
from $0.15^{\prime\prime}$ of arc separation to at least $2^\circ$
separation -- nearly 3 orders of magnitude. \cite{larson95} extended
this survey to smaller scales and found a break in the correlation at
0.04 pc. He suggested that smaller scales formed binary stars by
fragmentation in clumps, and larger scales formed hierarchical groups
by gas-related fragmentation processes, including turbulence.

Low mass x-ray stars in Gould's Belt (i.e., T Tauri stars) show a
hierarchical structure in all-sky maps \citep{guillout98}. This means
that large groupings of T Tauri stars contain smaller sub-groupings and
these contain even smaller sub-sub groupings. This is a much bigger
scale than the Taurus region studied by \cite{gomez93}. There has been
no formal correlation of the large-scale structure in x-ray stars yet.
The all-sky coverage suggests that the Sun is inside a hierarchically
clumped complex of young stars.

\section{The Cartwright \& Whitworth $Q$ parameter}

To study hierarchical structure in a different way, \cite{cw04}
introduced a parameter, $Q$. This is the ratio of the average
separation in a minimum spanning tree to the average 2-point
separation. For example, suppose there are 5 stars clustered together
in one region with a typical separation of 1 unit, and another 5 stars
clustered together in another region with a typical separation of 1
unit, and these two regions are separated by 10 units. Then the minimum
spanning tree has 4 separations of 1 unit in each region and 1
separation of 10 units (for the two closest stars among those two
regions), for an average of $(8\times1+1\times10)/(8+1)=2$ units
length. The average separation for all possible pairs is counted as
follows: there are 5 stars with separation from another star equal to
about 1 in each region, so that means 5 stars taken 2 at a time in each
region, or 10 pairs with a separation of 1 in each region, or 20 pairs
with this separation total, plus each star in one group has a
separation of 10 units from each star in the other group, which is
$5\times5$ separations of 10 units. The average is
$(20\times1+25\times10)/(20+25)=6$. The ratio of these is $Q=2/6=0.33$.
Smaller $Q$ means more subclumping because for multiple subgroups, the
mean 2-point separation has a lot of distances equal to the overall
size of the region, so the denominator of $Q$ is large, but the minimum
spanning tree has only a few distances comparable to the overall size
of the system, one for each subgroup, and then the numerator in $Q$ is
small.

\cite{bastian09} looked at the correlated properties of stars in the
LMC, using a compilation from \cite{zaritsky04}. There were about 2000
sources in each of several age ranges on the color-magnitude diagram.
Bastian et al. determined the zero-points and slopes of the two point
correlation function for each different age. They found that younger
regions have higher correlation slopes and greater correlation
amplitudes, which means more hierarchical substructure. Most of this
substructure is erased by 175 Myr. They also evaluated the \cite{cw04}
Q parameter and found a systematic decrease in $Q$ with decreasing age,
meaning more substructure for younger stars. \cite{gieles08} did the
same kind of correlation and $Q$ analysis for stars in the Small
Magellanic Cloud, and found the same general result.

\section{Hierarchies inside Clusters}

In a hierarchically structured region, the average density increases as
you go down the levels of the hierarchy to smaller and smaller scales.
If there are dense star-forming cores at the bottom of the hierarchy,
where the densities are largest and the sizes are smallest, then the
fractional mass in the form of these cores increases as their level is
approached.  This is because more and more interclump gas is removed
from the scale of interest as the densest substructures are approached.
The fractional mass of cores is proportional to the instantaneous
efficiency of star formation if the cores form stars. Therefore the
local efficiency of star formation in a hierarchical cloud increases as
the average density increases. The efficiency on the scale of a galaxy
where the average density is low is $\sim1$\%; on the scale of an OB
association it is $\sim5$\%, and in a cloud core where a bound cluster
forms, it is $\sim40$\%. Bound cluster formation requires a high
efficiency so there is a significant gravitating mass of stars
remaining after the gas leaves. It follows that in hierarchical clouds,
the probability of forming a bound cluster is automatically highest
where the density is highest. Star clusters are the inner bound regions
of a hierarchy of stellar and gaseous structures (Elmegreen 2008).

Outside the inner region, stars that form are not as likely to be bound
to each other after the gas leaves. Then there are loose stellar
groups, unbound OB subgroups, OB associations, and so on up to star
complexes. Flocculent spiral arms and giant spiral-arm clouds are the
largest scale on which gravitational instabilities drive the hierarchy
of cloud and star-formation structures.

The hierarchy of young stellar structure continues inside bound
clusters as well. \cite{smith05} found several levels of stellar
subclustering inside the rho-Ophiuchus cloud, and \cite{dahm05} found 4
subclusters with slightly different ages ($\pm1$ Myr) in NGC 2264.
Feigelson et al. (2009) observed X-rays from young stars in NGC 6334.
The x-ray maps are nearly complete to stars more massive than
$1\;M_\odot$ and their distribution is hierarchical, with clusters of
clusters inside this region. \cite{gutermuth05} studied azimuthal
profiles of clusters and found that they have intensity fluctuations
that are much larger than what would be expected from the randomness of
stellar positions; the stars are sub-clustered in a statistically
significant way. \cite{sanchez09} measured the fractal dimension and
hierarchical-$Q$ parameter for 16 Milky Way clusters, using the ratio
of cluster age to size as a measure of youth. They found that stars in
younger and larger clusters are more clumped than stars in older and
smaller clusters. Greater clumping means they have lower $Q$ and lower
fractal dimension. \cite{schmeja08} measured $Q$ for several young
clusters. For IC 348, NGC 1333, and Ophiuchus, $Q$ is lower (more
clumpy) for class 0/1 objects (young) than for class 2/3 objects (old).
Among four of the subclumps in Ophiuchus, $Q$ is lower and the region
is more gassy where class 0/1 dominates; $Q$ is also lower for class
0/1 alone than it is for class 2/3 in Ophiuchus.

Pretellar cores are spatially correlated too. \cite{johnstone00}
derived a power-law 2-point correlation function from $10^{3.8}$ AU to
$10^{4.6}$ AU for 850$\mu$m sources in Ophiuchus, which means they are
spatially correlated in a hierarchical fashion. \cite{johnstone01}
found a similar power-law from $10^{3.6}$ AU to $10^{5.1}$ AU for
850$\mu$m sources in Orion. Enoch et al. (2006) showed that 1.1 $\mu$m
pre-stellar clumps in Perseus have a power-law 2-point correlation
function from $10^{4.2}$ AU to $10^{5.4}$ AU. \cite{young06} found
similar correlated structure for pre-stellar cores from $10^{3.6}$ AU
to $10^5$ AU in Ophiuchus. These structures could go to larger scales,
but the surveys end there.

In summary, clusters form in the cores of the hierarchy of interstellar
structures and they are themselves the cores of the stellar hierarchy
that follows this gas. Presumably, this hierarchy comes from
self-gravity and turbulence. Gas structure continues to sub-stellar
scales. The densest regions, which are where individual stars form, are
always clustered into the next-densest regions. Stars form in the
densest regions, some independently and some with competition for gas,
and then they move around, possibly interact a little, and ultimately
mix together inside the next-lower density region. That mixture is the
cluster. More and more sub-clusters mix over time until the cloud
disrupts. Simulations of such hierarchical merging have been done by
many groups, such as \cite{bb06} and \cite{masch10}. Because of
hierarchical structure, the efficiency is automatically high on small
scales where the gas is dense.

\section{Clustered versus Non-clustered Star Formation}\label{sect:uc}

\cite{barba09} examined the giant star-forming region NGC 604 in M33
with NICMOS, finding mostly unclustered stars. \cite{maiz01}
categorized star formation regions according to three types: compact
clusters with weak halos, measuring $50\times50$ pc$^2$, compact
clusters with strong halos measuring $100\times100$ pc$^2$, and
hierarchical, but no clusters, called ``Scaled OB Associations''
(SOBAs), measuring $100\times100$ pc$^2$. Why do stars form in clusters
some of the time but not always?

The occurrence of bound clusters in star forming regions could depend
on many factors, but the pressure of the region relative to the average
pressure should be important. Higher pressure regions should produce
proportionally more clusters. Recall, that \cite{elias09} found a
higher clustering fraction for the high-pressure Orion region compared
to the low-pressure Sco-Cen region. One reason for a possible pressure
dependence was discussed in \cite{e08} and is reviewed here.

Turbulence produces a log-normal density probability distribution
function \citep{vaz94,price10}, and this corresponds to a log-normal
cumulative mass fraction $f_{\rm M} (>\rho)$, which is the fraction of
the gas mass with a density larger than the value $\rho$. This is a
monotonically decreasing function of $\rho$. If the densest clumps have
a density $\rho_{\rm c}$, and the star formation rate per unit volume
is the dynamical rate for all densities with an efficiency (star-to-gas
mass fraction) that depends on density, i.e.,
$SFR=\epsilon(\rho)\rho\left(G\rho\right)^{1/2}$, then the mass
fraction of the densest clumps inside a region of average density
$\rho$ is
\begin{equation}
\epsilon(\rho)=\epsilon_{\rm c}\left(\rho_{\rm
c}/\rho\right)^{1/2}\left[ f_{\rm M}(>\rho_{\rm c})/ f_{\rm
M}(>\rho)\right].\end{equation} This function $\epsilon(\rho)$
increases with $\rho$ for intermediate to high density. If stars form
in the densest clumps with local efficiency $\epsilon_{\rm c}$
($\sim0.3$), then $\epsilon(\rho)$ is the efficiency of star formation,
i.e., the mass fraction going into stars for each average density.
Bound clusters form where the efficiency is highest, and this is where
the average density is highest.  If we consider the density where
$\epsilon(\rho)$ exceeds a certain minimum value for a bound cluster,
then most star formation at this density or larger ends up in bound
clusters.

Note that the observation of cluster boundedness appears to be
independent of cluster mass and therefore independent of the presence
of OB stars in the cluster. Bound clusters with highly disruptive OB
stars form in dense cloud cores, just like clusters without these
stars. The efficiency of star formation is therefore not related in any
obvious way to the presence or lack of disruptive stars. The
implication is that essentially all of the stars in a cluster form
before OB-star disruption occurs. Perhaps the highly embedded nature of
OB star formation, in ultracompact HII regions, for example, shields
the rest of the cloud core from disruption for a long enough time to
allow the lower mass cores to collapse into stars.

The pressure dependence for cluster boundedness arises in the theory of
Elme\-green (2008) because at a fixed density for star formation, the
slope of the $f_{\rm M}(>\rho)$ curve decreases for higher average
density (the log-normal shifts to higher density), and the slope also
decreases for higher Mach number because the log-normal gets broader.
With a shallower slope at the density of star formation, the density
where $\epsilon(\rho)$ exceeds the limit for bound cluster formation
decreases, and the fraction of the mass exceeding this density
increases. Thus forming a bound cluster happens at a lower density
relative to the threshold for star formation when the pressure is high.
A higher fraction of the gas then goes into bound clusters. The
qualitative nature of this conclusion is independent of the details of
the density pdf.

This discussion of cluster formation in a hierarchical medium assumes
that the gas structure is already in place when star formation begins,
and then the densest clumps, which are initially clustered together,
form stars. The discussion can be made more dynamical if we consider
that the denser regions fragment faster. This is what happens during a
collapse simulation: the dense regions make more subcondensations and
the low density regions collapse on to the dense regions. The result is
a clustering of dense sub-condensations and a hierarchical clustering
of stars.

\section{Cluster Mass Functions}\label{mf}

The cluster mass function is a power law and it is natural to look for
explanations of this that are related to the other power laws in star
formation, including the hierarchical structure.  If we imagine a cloud
divided hierarchically into clumps and sub-clumps, then the mass
distribution function of the nodes in this hierarchy is $dN/d\log
M\sim1/M$, or $dn/dM\sim1/M^2$, because there is an equal total mass in
all levels ($MN[M]d\log M=$ constant). This is the same as the mass
function for star clusters. Also in such a hierarchy, the probability
that a mass between $M$ and $2M$ is selected is proportional to $1/M$,
as given by the number of levels and clouds at those levels having a
mass in that range.

Cluster mass functions typically are a power law with a slope equal to
this value, $dn/dM\sim M^{-\beta}$ for $\beta\sim2$. This slope was
found by \cite{batt94} for the solar neighborhood and \cite{eef97} for
the LMC, where the clusters were subdivided according to age.  A second
study of LMC clusters \citep{hunter03} found the same slope for
discrete cluster age intervals. It is important to consider clusters
within a narrow age interval because older clusters are dimmer and the
selection effects for clusters depend on their age. \cite{zf99} found
$\beta=1.95\pm0.03$ for young clusters in the Antenna galaxy, and
$\beta=2.00\pm0.08$ for old clusters. \cite{degrijs06} looked at the
LMC again and found $\beta=1.85\pm0.05$ for various age intervals.
\cite{degrijs03} found similar results in two other galaxies:
$\beta=2.04\pm0.23$ for NGC 3310 and $\beta=1.96\pm0.15$ for NGC 6745.

The HII region luminosity function is about the same as the cluster
mass function, having a slope of around $-2$ for linear intervals of
luminosity. The first large study was by \cite{kennicutt89}. Many other
surveys have obtained about the same result \citep[e.g.,][]{banfi93}.
\cite{bradley06} included 53 spiral galaxies and got a steeper slope at
$\log L> 38.6$ (for $L$ in erg s$^{-1}$), suggesting that larger HII
regions were density bounded, and they also got a steep fall-off at
$\log L>40$, suggesting an upper limit for cluster mass. The same
general power law for HII regions has been obtained in detailed studies
of individual galaxies (e.g., NGC 3389: Abdel-Hamid et al. (2003); M81:
Lin et al. 2003; NGC 1569: Buckalew \& Kobulnicky 2006; NGC 6384:
Hakobyan et al. 2007; the Milky Way: Paladini et al. 2009).

There is growing evidence for an upper mass cutoff in the cluster mass
function. In \cite{gieles06a,gieles06b}, mass functions in M51 were fit
to a double power law, i.e., with an increased slope at higher mass, or
to a power law with $\beta=2$ throughout and an upper mass cutoff of
around $10^5\;M_\odot$. A power law with an exponential cutoff is a
Schechter function, $dN/dM=M^{-\beta}\exp(-M/M_{\rm c})$ for cutoff
mass $M_{\rm c}$.

For several local galaxies, \cite{larsen09} fit the brightest cluster
and the 5th brightest cluster with a mass function having a cutoff.
Larsen found that rich and poor spirals have about the same cluster
mass functions, both with a cutoff, and that the cutoffs are
independent of position in a galaxy. The origin of an upper mass limit
for clustering is not known.

\section{Summary}

Gas is hierarchical in space and time, presumably because the gas is
compressed by turbulent motions in a scale-free fashion. The
self-gravitational force is scale free also at masses far above the
thermal Jeans mass ($\sim1\;M_\odot$). For a typical relationship
between velocity dispersion and size that scales as $\sigma\propto
R^{1/2}$, clouds of all masses at constant pressure have the same
degree of gravitational self-binding.

Hierarchical cloud structure means that stars form in hierarchical
patterns, and it follows then that the efficiency of star formation
($M_{\rm stars}/M_{\rm total}$) increases with the average density.
Bound star clusters, which require a high efficiency, therefore form at
high density. This explains at a very fundamental level why bound
clusters form in the first place. Variations in the fraction of star
formation that goes into bound clusters may be explained in the same
way, with pressure playing an important role.

Hierarchical structure ensures that the clusters start with a mass
function that is a power law with a slope close to $-2$. There could be
an upper mass cut off.

%%-----------------------------
%%      your bibliography
%%-----------------------------

\end{document}